\documentclass[aps,prl,twocolumn,superscriptaddress,letterpaper]{revtex4-1}

\usepackage{graphicx}
\usepackage{bm}
\usepackage{amssymb}
\usepackage{color}
\usepackage{amsmath}
\usepackage{ulem}

\newcommand{\be}{\begin{equation}}
\newcommand{\ee}{\end{equation}}

\begin{document}

\title{Non-Hermitian Dirac Cones}

\author{Haoran Xue}
\affiliation{Division of Physics and Applied Physics, School of Physical and Mathematical Sciences, Nanyang Technological University,
Singapore 637371, Singapore}

\author{Qiang Wang}
\affiliation{Division of Physics and Applied Physics, School of Physical and Mathematical Sciences, Nanyang Technological University,
Singapore 637371, Singapore}

\author{Baile Zhang}
\email{blzhang@ntu.edu.sg}
\affiliation{Division of Physics and Applied Physics, School of Physical and Mathematical Sciences, Nanyang Technological University,
Singapore 637371, Singapore}
\affiliation{Centre for Disruptive Photonic Technologies, Nanyang Technological University, Singapore, 637371, Singapore}

\author{Y. D. Chong}
\email{yidong@ntu.edu.sg}
\affiliation{Division of Physics and Applied Physics, School of Physical and Mathematical Sciences, Nanyang Technological University,
Singapore 637371, Singapore}
\affiliation{Centre for Disruptive Photonic Technologies, Nanyang Technological University, Singapore, 637371, Singapore}

\begin{abstract}
Non-Hermitian systems, which contain gain or loss, commonly host exceptional point degeneracies rather than the diabolic points found in Hermitian systems.  We present a class of non-Hermitian lattice models with symmetry-stabilized diabolic points, such as Dirac or Weyl points.  They exhibit non-Hermiticity-induced phenomena previously existing in the Hermitian regime, including topological phase transitions, Landau levels induced by pseudo-magnetic fields, and Fermi arc surface states.  These behaviors are controllable via gain and loss, with promising applications in tunable active topological devices.
\end{abstract}

\maketitle

\textit{Introduction.}---If a Hamiltonian is Hermitian, its eigenstates form an orthogonal basis, and eigenvalue degeneracies are diabolic points.  The presence of diabolic points, such as Dirac points in two-dimensional (2D) lattices, has important implications \cite{neto2009,armitage2018,bradlyn2016}: their dynamics is described by elementary equations like the 2D Dirac equation \cite{Novoselov2005,xu2015,lu2015}, giving rise to nontrivial topological features; for instance, Dirac points play a key role in the transitions between topologically distinct 2D insulator phases \cite{haldane1988, Bansil2016}.  Recently, there has also been a great deal of interest in degeneracies found in non-Hermitian systems, which are systems with gain and/or loss \cite{Bender2007, Moiseyev2011}.  The eigenvectors of a non-Hermitian Hamiltonian are not guaranteed to be orthogonal, and generally the degeneracies are exceptional points (EPs), meaning that the eigenvectors and not just the eigenvalues are degenerate \cite{kato1966}.  This includes parity/time-reversal (PT) symmetric Hamiltonians, which can exhibit real eigenvalues with non-orthogonal eigenvectors \cite{Bender1998, Feng2017}.  EPs give rise to many interesting physical effects, with possible applications in waveguide mode conversion, optical sensing, and more \cite{Bender1998, Feng2017, el2018, miri2019, lin2011, dembowski2001, wiersig2014, Leykam2017, kawabata2019, yang2019}.  The relationship between diabolic points and EPs have also come under scrutiny.  Diabolic points have been found to split into EPs under non-Hermitian perturbations \cite{zhen2015,zhou2018,xu2017,cerjan2018,Cerjan2019,budich2019,okugawa2019,zhou2019}; for example, a Dirac point in a 2D Hermitian lattice can turn into an exceptional ring \cite{zhen2015} or two EPs \cite{zhou2018}, while a Weyl point in a three-dimensional (3D) lattice can become a ring of EPs \cite{Cerjan2019}. However, little attention has been paid to the existence of diabolic points in non-Hermitian systems. There seem to be no fundamental constraints preventing non-Hermitian systems from exhibiting diabolic points, but what are the conditions to achieve this, and what are the physical consequences?

This Letter proposes a mechanism for non-Hermitian systems to support symmetry-stabilized diabolic points. We present a class of Hamiltonians that are not Hermitian, but obey a set of symmetries ensuring the existence of pairs of real eigenvalues with orthogonal eigenvectors. We then describe a 2D non-Hermitian lattice with Dirac points.  In this model, gain and loss can drive a topological transition between Chern insulator and conventional insulator phases \cite{Bansil2016}, or generate a pseudo-magnetic field inducing the formation of Landau levels \cite{Guinea2010, Rechtsman2013}. Extending the model to 3D, we demonstrate non-Hermitian Weyl points with complex-valued Fermi arcs \cite{wan2011}.

It should be noted that this work differs from the recent efforts \cite{Leykam2017, shen2018, yao2018, yao2018a, kunst2018, gong2018, zhou2019a, kawabata2019a} to extend the concepts of band geometry and topology to the non-Hermitian regime by formulating new topological invariants, topological classifications, bulk-boundary correspondences, etc.  The non-Hermitian models studied here acquire their interesting features not from novel topological principles, but from diabolic (Dirac or Weyl) points that act similarly (but not identically) to those in Hermitian models.  This points to the attractive possibility of realizing phenomena like as topological phase transitions or pseudo-magnetic fields in active devices in which the non-Hermitian parameters are tunable, such as photonic devices with actively controlled gain or loss.

\textit{Non-Hermitian symmetries.}---We first define the following $4\times4$ matrices:
\begin{equation}
  \Sigma_\mu = \begin{bmatrix}0 & \sigma_\mu \\ \sigma_\mu & 0
  \end{bmatrix}, \;\;\;\mu = 0, 1, 2, 3.
  \label{sigma-matrices}
\end{equation}
Here, $\sigma_0$ denotes the $2\times2$ identity matrix and $\sigma_j$ denotes the Pauli matrices for $j = 1, 2, 3$.

Let $H$ be a $4\times 4$ matrix, which needs not to be Hermitian, that satisfies (i) the pseudo-Hermiticity condition
\begin{equation}
  \Sigma_0 H \Sigma_0 = H^\dagger, \label{pseudo-Herm}
\end{equation}
and (ii) the anti-PT symmetry condition
\begin{equation}
  \big\{H, \Sigma_3 \Sigma_1 T \big\} = 0, \label{anti-PT}
\end{equation}
where $T$ is the complex conjugation operator.

The first condition, Eq.~\eqref{pseudo-Herm}, implies that the eigenvalues of $H$ are real or appear in complex conjugate pairs \cite{Mostafazadeh2002}.  The second condition, Eq.~\eqref{anti-PT}, implies that the eigenvectors form orthogonal pairs; to see this, note that if $|\psi_+\rangle$ is an eigenvector of $H$ with eigenvalue $E_+$, then $|\psi_-\rangle = \Sigma_1\Sigma_3 T |\psi\rangle$ is an eigenvector with eigenvalue $- E^*$.  Then $|\psi_+\rangle$ and $|\psi_-\rangle$ can be shown to be orthogonal \cite{SM}:
\begin{equation}
  \langle \psi_+ | \psi_- \rangle
  = \sum_{n=1}^4 \left(\psi_{+}^n\right)^* \psi_-^n = 0.
\end{equation}
The proof of this uses the specific form of $\Sigma_1 \Sigma_3$, and the fact that $\langle\varphi| \sigma_2 T |\varphi\rangle = 0$ for any two-component $|\varphi\rangle$.

With both symmetries simultaneously present, the eigenvalues of $H$ must satisfy one of two possibilities: either they form the set $\{z, z^*, -z, -z^*\}$ for some non-real $z$, or they form two real pairs $\{E_1, -E_1\}$ and $\{E_2, -E_2\}$, where the eigenvectors within each pair are orthogonal (but eigenvectors in different pairs are generally not orthogonal).  We refer to these distinct cases as ``symmetry-broken'' and ``symmetry-unbroken'' respectively.  As described below, a non-Hermitian system in the symmetry-unbroken regime can exhibit Dirac points and other phenomena previously existing only in Hermitian systems.  

A matrix satisfying Eqs.~\eqref{pseudo-Herm} and \eqref{anti-PT} has the form
\begin{equation}
  H = \begin{bmatrix}\mathcal{W} & \mathcal{V}_+ \\
    \mathcal{V}_- & \mathcal{W}^\dagger \end{bmatrix},
  \;\; \mathcal{V}_\pm = \mathcal{V}_\pm^\dagger,
  \label{H-form}
\end{equation}
where $\mathcal{W}$, $\mathcal{V}_+$, and $\mathcal{V}_-$ are $2\times2$ sub-matrices of the form
\begin{equation}
  \mathcal{W}
  = \begin{bmatrix}a & b \\ b^* & -a^* \end{bmatrix},
  \;\;\;
  \mathcal{V}_\pm
  = \begin{bmatrix}\lambda_\pm & c_\pm \\ c_\pm^* & -\lambda_\pm \end{bmatrix},
  \label{submatrix-conditions}
\end{equation}
for $a, b,c_\pm \in \mathbb{C}$ and $\lambda_\pm \in \mathbb{R}$.

\begin{figure}
  \centering
  \includegraphics[width=\columnwidth]{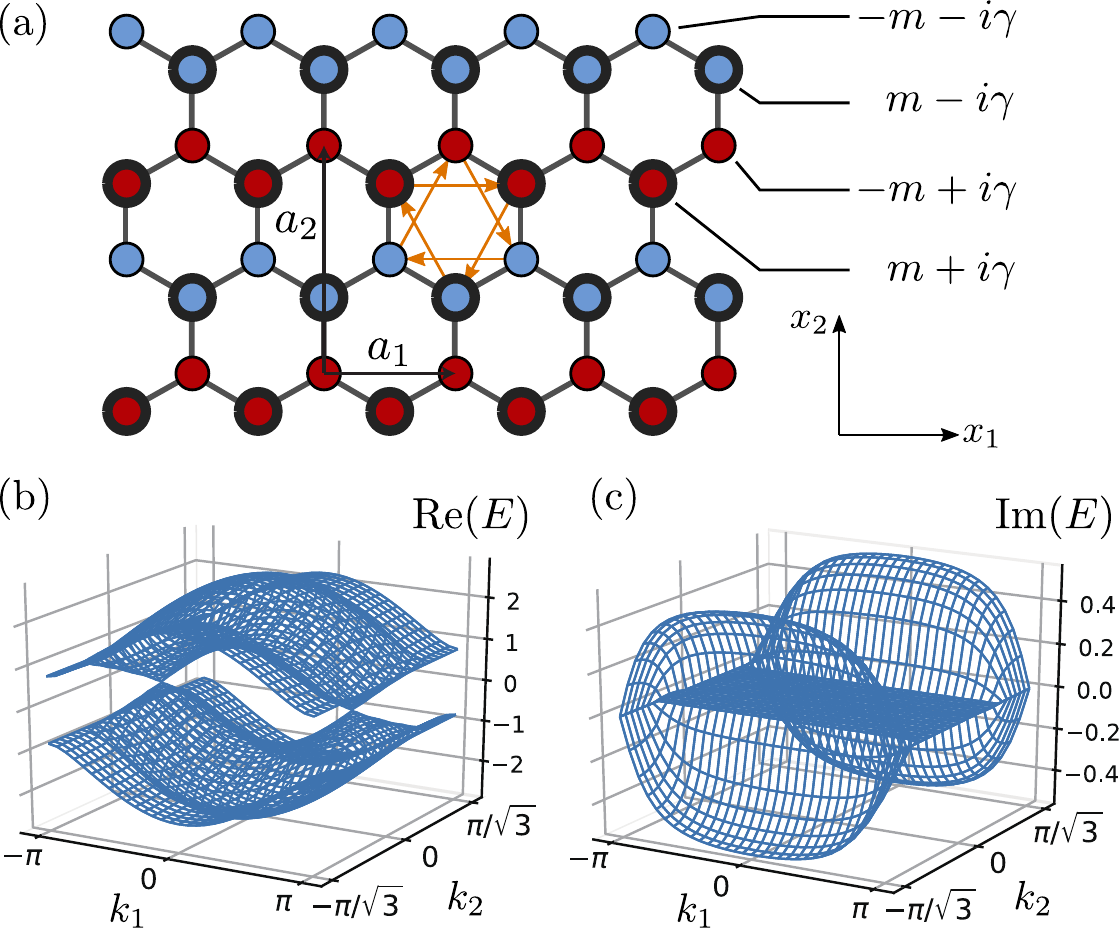}
  \caption{(a) Schematic of the non-Hermitian lattice.  Thick (thin) circle outlines indicate positive (negative) real parts of the on-site mass terms, $\pm m$, while red (blue) colors indicate positive (negative) imaginary parts, $\pm i \gamma$.  The nearest-neighbor coupling is $t_1 = 1$. Orange arrows indicate the next-nearest-neighbor couplings of $\mp it_2$ along (opposite to) the arrow; for clarity, only couplings in one hexagon are depicted.  Black arrows show the elementary lattice vectors $a_1$ and $a_2$. (b)--(c) The complex bandstructure for $m = 0$ and $\gamma=0.6$.  In the symmetry-unbroken domain, there are two Dirac cones with real energies and orthogonal eigenvectors.}
  \label{fig:lattice}
\end{figure}

\textit{Lattice model.}---Consider the honeycomb lattice shown in Fig.~\ref{fig:lattice}(a).  The lattice sites have complex on-site mass terms; the real parts $\pm m$ are indicated by thick and thin outlines, and the imaginary parts $\pm \gamma$ (i.e., on-site gain or loss) are indicated by red and blue colors.  Note that the real and imaginary terms are distributed differently.  Let the nearest-neighbor inter-site couplings be $t_1=1$, and take Haldane-type next-nearest-neighbor couplings with magnitude $t_2$ and phase $-\pi/2$.  For $\gamma = 0$, this reduces to the Haldane model \cite{haldane1988}.  For $\gamma \ne 0$, the lattice is non-Hermitian and each unit cell has four sites, with lattice vectors $a_1 = [1,0]$ and $a_2 = [0,\sqrt{3}]$.

The $k$-space Hamiltonian $H(k)$ satisfies Eqs.~\eqref{H-form}--\eqref{submatrix-conditions}, where the submatrix parameters in \eqref{submatrix-conditions} are
\begin{align}
  \begin{aligned}
    a &= m + i \gamma + 2t_2 \sin k_1 \\
    b &= 2\cos\left(k_1/2\right)\, \exp\big[ik_2/(2\sqrt{3})\big] \\
    \lambda_\pm &= -4t_2 \sin(k_1/2) \cos(\sqrt{3}\,k_2/2) \\
    c_\pm &= \exp\big(\!-ik_2/\sqrt{3}\,\big).
  \end{aligned}
\end{align}
Thus, Eqs.~\eqref{pseudo-Herm} and \eqref{anti-PT} hold for all $k$.  If the next-nearest-neighbor couplings are nonzero and have phases other than $\pm\pi/2$, $H(k)$ would not satisfy the symmetries.

The spectrum of $H(k)$ can be derived analytically for $m = t_2 = 0$ \cite{SM}.  For $|\gamma| < 1$, all four eigenvalues are real for $|k_2| < \cos^{-1}\left(2\gamma^2-1\right)/\sqrt{3}$, and in the symmetry-unbroken domain there are $E = 0$ band degeneracies at
\begin{equation}
  K^\tau = \begin{pmatrix}- 2\tau\theta \\ 0 \end{pmatrix},
  \;\; \mathrm{where}\;
  \begin{cases} \tau &= \pm 1, \\ \cos2\theta &= (-1-\gamma^2)/2.
  \end{cases}
  \label{Kpoints}
\end{equation}
Fig.~\ref{fig:lattice}(b)--(c) shows the band structure for $m = t_2 = 0$ and $\gamma = 0.6$, with the degeneracy points clearly visible.

\begin{figure*}
  \centering
  \includegraphics[width=\textwidth]{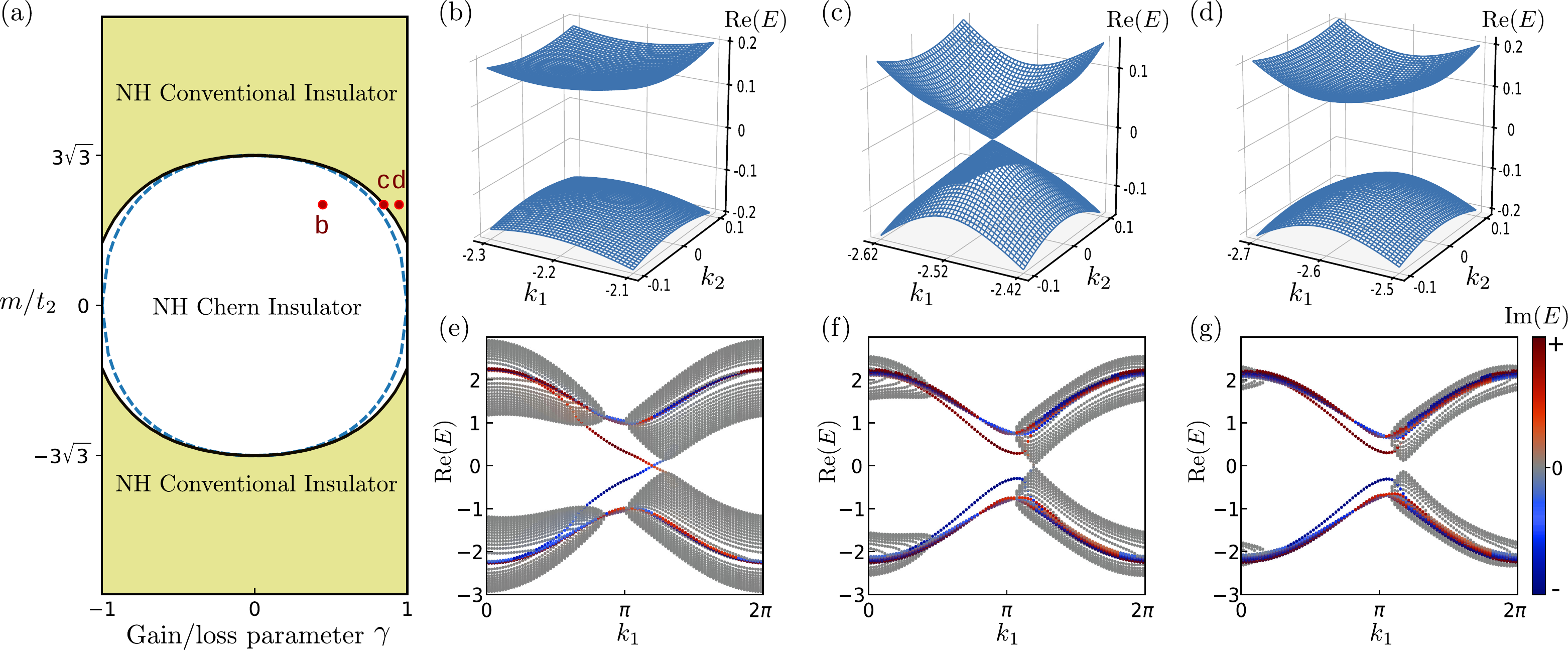}
  \caption{(a) Phase diagram of the 2D non-Hermitian lattice, featuring a non-Hermitian Chern insulator phase (white) and a non-Hermitian conventional insulator phase (yellow).  Black curves are phase boundaries computed by varying $m$ and $\gamma$ for fixed $t_2 = 0.1$ and searching numerically for band degeneracies.  Blue dashes are analytical phase boundaries given by Eq.~\eqref{phase-transition}.  Red dots indicate the points corresponding to the (b)--(d) subplots.  (b)--(d) Bulk band diagrams near one of the $K^\tau$ points.  The lattice parameters are $t_2 = 0.1$, $m = 0.35$, and $\gamma = 0.846 +\delta\gamma$ where $\delta\gamma = -0.4$ for the non-Hermitian Chern insulator (b), $\delta\gamma = 0$ at the phase boundary (c), and $\delta\gamma = 0.1$ for the non-Hermitian conventional insulator (d).  All depicted eigenvalues have zero imaginary part.  (e)--(g) Band diagrams for a strip that is infinite in the $x_1$ direction and 25 cells wide in the $x_2$ direction, using the parameters corresponding to (b)--(d) respectively.  Colors represent the imaginary part of the energy eigenvalues.}
  \label{fig:phasediag}
\end{figure*}

One can verify numerically that there are two orthogonal eigenstates at each degeneracy point, so these are diabolic points and not EPs despite $H(k)$ being non-Hermitian.  In Fig.~\ref{fig:lattice}(b)--(c), we see that the spectrum is linear around each degeneracy point, indicating that they are 2D Dirac points.  To prove this, let $q = (q_1, q_2)$ be the $k$-space displacement relative to $K^\tau$, let $|\psi\rangle = [|\varphi_+\rangle, |\varphi_-\rangle]$ be an eigenstate with energy $E$, and let $E$, $m$, and $t_2$ be of order $q$.  To first order, $|\varphi_\pm\rangle$ are found to be governed by 2D Dirac Hamiltonians \cite{SM}:
\begin{equation}
  v \Big(\tau \eta_\gamma \sigma_1 q_1 - \sigma_2 q_2
  + M \sigma_3 \Big) U_\pm |\varphi_\pm\rangle = E \, U_\pm |\varphi_\pm\rangle
  + \cdots
  \label{Dirac}
\end{equation}
Here $v = \sqrt{3}/2$ is the Dirac velocity, $\eta_\gamma = 2\sin\theta/\sqrt{3}$ is an anisotropy parameter that goes to 1 when $\gamma \rightarrow 0$, and
\begin{align}
  U_\pm &= \exp\left[\pm \frac{i}{2} (\sin^{-1}\gamma)\, \sigma_1\right] \\
  M &= \frac{m - 6\tau t_2 \sin2\theta}{2\cos\theta}. \label{Mpm}
\end{align}
This result only applies to the two bands involved in the degeneracy point; the other two do not satisfy $E \sim O(q)$, so Eq.~\eqref{Dirac} does not apply.

\textit{Non-Hermitian gapped phases.}---For $t_2 \ne 0$, the model exhibits two types of gapped phases.  If the term proportional to $\tau t_2$ in Eq.~\eqref{Mpm} dominates the term proportional to $m$, the Dirac cones have opposite mass, similar to the Chern insulator phase of the Haldane model \cite{haldane1988}.  If the reverse is true, the Dirac cones have the same mass, like the conventional insulator phase of the Haldane model.  The phase transition is thus predicted to occur at
\begin{equation}
  |m| = 3|t_2| \sqrt{(3+\gamma^2)(1-\gamma^2)}.
  \label{phase-transition}
\end{equation}
At the transition point, there is an unpaired Dirac cone at one of the $K^\tau$ points, similar to the transition described in the Haldane model\cite{haldane1988}.

The phase diagram is shown in Fig.~\ref{fig:phasediag}(a).  The phase boundaries predicted by Eq.~\eqref{phase-transition} agree well with those found by numerically searching band degeneracies; the discrepancies become smaller as $t_2$ is further reduced.  Moreover, the shape of the transition curve points to the interesting possibility of driving a phase transition entirely via gain and loss.  As indicated by the points labeled b--d in Fig.~\ref{fig:phasediag}(a), we can fix nonzero values of $m$ and $t_2$ and increase $\gamma$ from zero, and thereby change the system from a non-Hermitian Chern insulator (which reduces to a Hermitian Chern insulator for $\gamma = 0$) into a non-Hermitian conventional insulator [Fig.~\ref{fig:phasediag}(b)--(d)].
  
The two gapped bulk phases are accompanied by edge state behaviors similar to Hermitian Chern and conventional insulators, as seen in Fig.~\ref{fig:phasediag}(e)--(g).  The non-Hermitian Chern insulator hosts edge states that span the gap.  These edge states do not appear to be linked to a rigorous non-Hermitian bulk-edge correspondence principle \cite{Lee2016, Leykam2017}; rather, they are a consequence of the fact that the non-Hermitian $k$-space Hamiltonian has effective 2D Dirac solutions, to which the standard Hermitian bulk-edge correspondence applies.  The truncation of the real-space lattice breaks the symmetries \eqref{pseudo-Herm}--\eqref{anti-PT}, spoiling the reality and orthogonality properties of the eigenstates.  As a consequence, the edge state energies acquire substantial imaginary parts; these values depend on the choice of boundary termination, and in the case of Fig.~\ref{fig:phasediag}(e)--(g) range from around -0.9 to 0.9.  For most of the bulk states, the energies are almost real.

\begin{figure}
\centering
\includegraphics[width=\columnwidth]{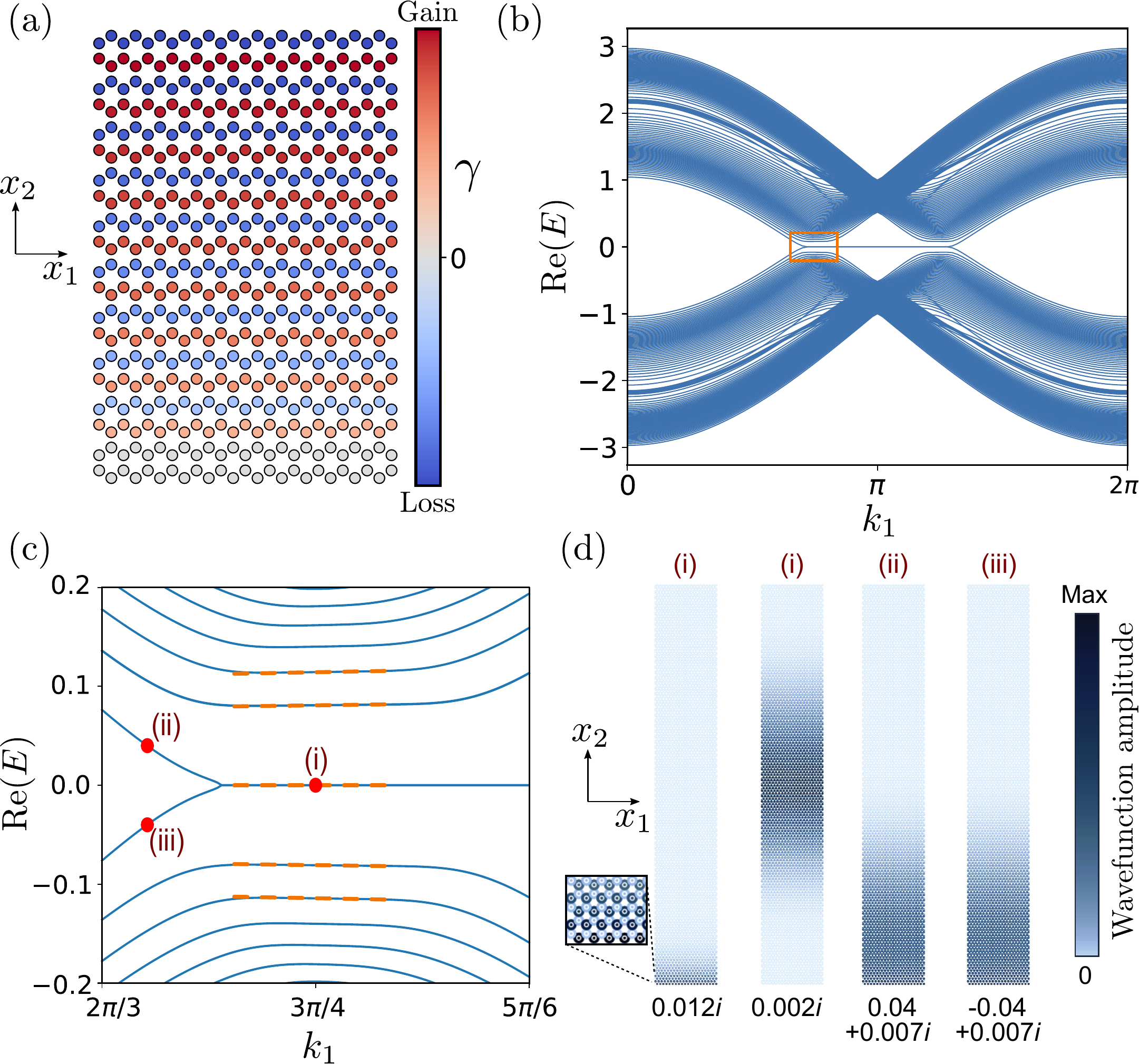}
\caption{(a) Schematic of a semi-infinite lattice with gradually varying gain and loss, distributed so that $\theta = \theta_0 + \beta x_2$.  The lattice is infinite in the $x_1$ direction.  (b) Real part of the band diagram for a semi-infinite lattice 300 sites wide in the $x_2$ direction, with $\beta =0.002$.  (c) Close-up view of the band diagram near one of the projected $K^\tau$ points, showing the formation of Landau levels.  Horizontal dashes indicate the energies for the zeroth Landau level and first two pairs of nonzero Landau levels, derived using a continuum Dirac theory with a pseudo-magnetic field.  (d) Wavefunction amplitude distributions for the four eigenstates indicated by (i)--(iii) in (c).  The numbers below indicate the energy eigenvalues.}
\label{fig:strain}
\end{figure}

\textit{Non-Hermitian pseudo-magnetic fields.}---The non-Hermitian Dirac cones can experience pseudo-magnetic fields similar to those found in Hermitian lattices \cite{Guinea2010, Rechtsman2013}, except that these pseudo-magnetic fields can be induced by gain/loss engineering rather than strain engineering.  For $m = t_2 = 0$, the Dirac cones are ungapped; since their $k$-space positions depend on $\gamma$ [Eq.~\eqref{Kpoints}], a spatially non-uniform variation in $\gamma$ acts as a gauge field.  By analogy, in Hermitian graphene-like lattices a spatially uniform change in the coupling terms shifts the Dirac points in $k$-space, and a non-uniform variation acts as a valley-specific gauge field that can induce Landau levels \cite{Guinea2010, Rechtsman2013}.

Fig.~\ref{fig:strain}(a) shows a schematic of a lattice that is infinite in the $x_1$ direction and finite along $x_2$.  Since the $k$-space displacement of the Dirac point is proportional to the $\theta$ variable introduced in Eq.~\eqref{Kpoints}, we choose to vary the gain/loss parameter $\gamma$ so that $\theta(\gamma(x_2)) = \theta_0 + \beta x_2$.  A theoretical analysis \cite{SM} shows that the resulting system hosts a zeroth Landau level (a flat band at $E = 0$), and bands similar to nonzero Landau levels.
The numerically obtained band diagram is shown in Fig.~\ref{fig:strain}(b)--(c).  The zeroth Landau level is clearly present, and the energies of the nonzero bands are also close to theoretical predictions. In Fig.~\ref{fig:strain}(d) we plot the wavefunction amplitude distributions for four eigenstates, labelled by (i)--(iii) in Fig.~\ref{fig:strain}(c).  The eigenstates at (i) are two-fold degenerate, and consist of a bulk mode and an edge mode, as seen in the first two plots of Fig.~\ref{fig:strain}(d).  The eigenstates at (ii) and (iii) are edge modes connected to the zeroth Landau level.  We emphasize that this phenomenon is generated by a spatial variation in the gain/loss, with no strain engineering \cite{schomerus2013,zhang2019} or real magnetic field.  

\begin{figure}
\centering
\includegraphics[width=\columnwidth]{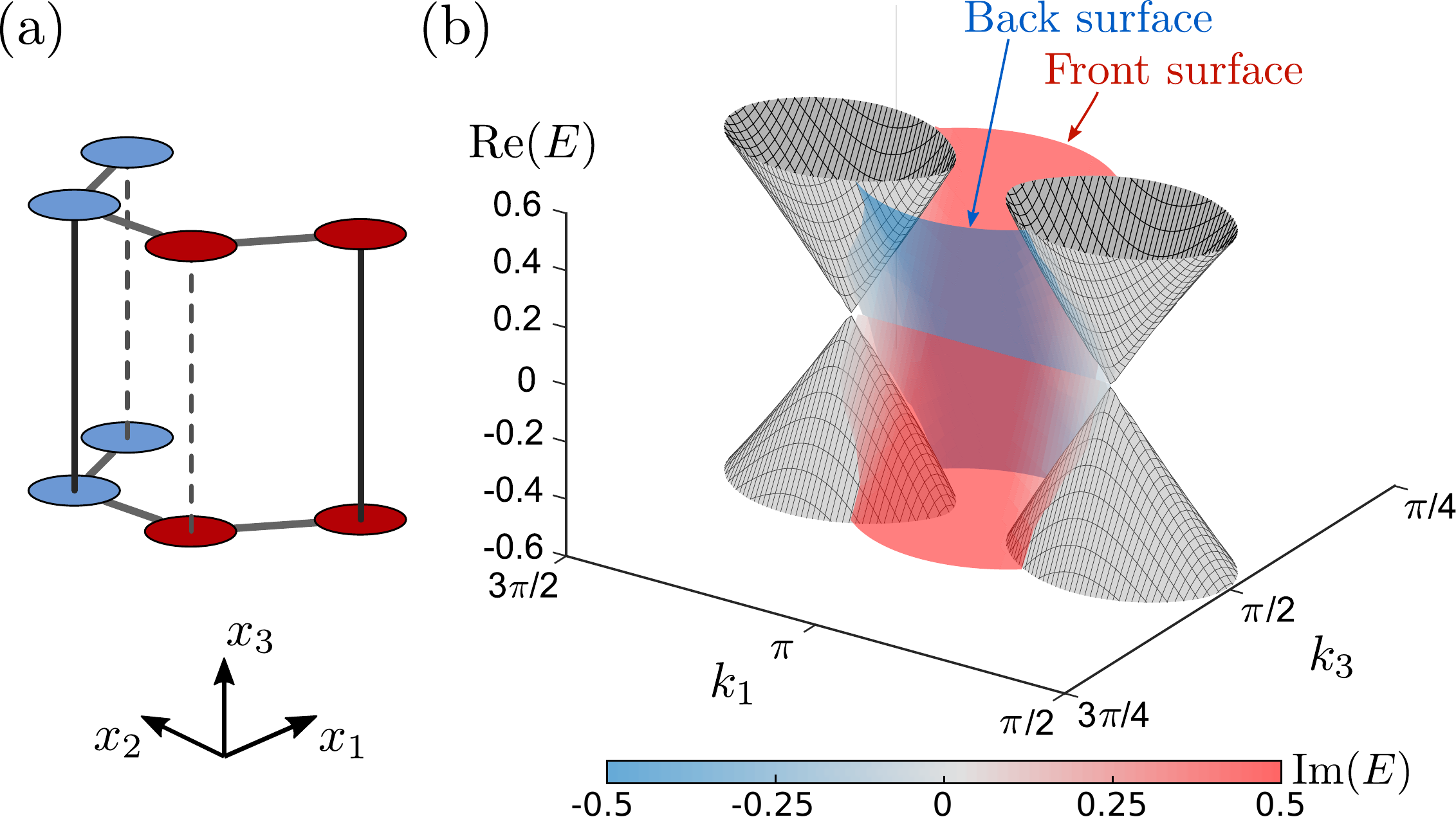}
\caption{(a) Schematic of a 3D non-Hermitian lattice hosting Weyl points. Solid (dashed) lines indicate couplings with positive (negative) signs $\pm t_3$. (b) Surface dispersion for a slab geometry with periodic boundary conditions along $x_1$ and $x_3$ diractions and open boundary condition along $x_2$.  The lattice parameters are $t_1 = t_3 = 1$, $t_2 = m = 0$, and $\gamma = 0.6$.}
\label{fig:weyl}
\end{figure}

\textit{Non-Hermitian Weyl points}---The present framework for generating diabolic points in non-Hermitian systems is not limited to Dirac points in 2D lattices.  As an example, we formulate a non-Hermitian 3D lattice model exhibiting Weyl points, the simplest 3D diabolic points.

The lattice is shown schematically in Fig.~\ref{fig:weyl}(a).  It is constructed by stacking the 2D lattice in Fig.~\ref{fig:lattice}(a) with $m = t_2 = 0$.  In the $x_3$ direction, the interlayer couplings are either postive or negative, as indicated by solid or dashed lines.  The corresponding lattice Hamiltonian still satisfies Eqs.~\eqref{H-form}--\eqref{submatrix-conditions} since for each $k_3$ the interlayer couplings act like the $m$ parameter.  In the Hermitian limit ($\gamma=0$), the system is a Weyl semimetal \cite{xiao2015}.  As $\gamma$ is varied, the Weyl points shift in momentum space (this raises the possibility of chiral Landau levels \cite{pikulin2016} realized by gain/loss engineering). In Fig.~\ref{fig:weyl}(b), we observe the formation of complex-valued Fermi arcs, whose real parts connect the projections of the real-valued bulk Weyl cones.

\textit{Discussion.}---We have shown that diabolic points can occur in non-Hermitian systems by using symmetry constraints to enforce eigenstate orthogonality.  The models we have studied are not the only ones that can achieve such outcomes; in fact, it is possible to formulate other non-Hermitian models that exhibit Dirac points without the symmetries \eqref{pseudo-Herm} and \eqref{anti-PT}.  It would be desirable to find a general description of all such non-Hermitian models.  It may also be interesting to examine other types of non-Hermitian diabolic points, such as quadratic band degeneracies \cite{Chong2008}, type-II Weyl points \cite{Noh2017}, higher order Weyl points \cite{bradlyn2016,Chen2016}, nodal lines \cite{Chen2016a}, etc.

Laser-written optical waveguide arrays \cite{Szameit2010} are a promising platform for realizing these non-Hermitian models.  Previously, it has been shown that waveguide losses in these arrays can be individually customized (to realize non-Hermitian bulk topological transitions \cite{Zeuner2015} or Weyl exceptional rings \cite{Cerjan2019}), while T can be effectively broken by twisting the waveguides \cite{Rechtsman2013a, Noh2017}.  It would also be highly desirable to realize these phenomena in active nanophotonic platforms such as resonator arrays \cite{Hafezi2011, Leykam2018, Mittal2019,zhao2019} with actively tunable gain or loss.  A longstanding obstacle to applying strain engineering concepts \cite{Guinea2010} to photonics is that mechanically deforming photonic devices is usually impractical.  The present scheme allows for using active gain or loss to generate a pseudo-magnetic field to alter the photonic density of states on demand.

We are grateful to M.~C.~Rechtsman for helpful discussions.  This work was supported by the Singapore MOE Academic Research Fund Tier 3 Grant MOE2016-T3-1-006, Tier 1 Grants RG187/18 and RG174/16(S), and Tier 2 Grant MOE 2018-T2-1-022(S).

\bibliography{references}

\clearpage

\begin{widetext}

\makeatletter 
\renewcommand{\theequation}{S\arabic{equation}}
\makeatother
\setcounter{equation}{0}

\makeatletter 
\renewcommand{\thefigure}{S\@arabic\c@figure}
\makeatother
\setcounter{figure}{0}

\begin{center}
{\Large Supplemental Material}
\end{center}

\section{Consequences of the Non-Hermitian Symmetries}

Here, we provide a more detailed discussion of the consequences of the symmetries described in the main text,
\begin{align}
  \Sigma_0 H \Sigma_0 &= H^\dagger, \label{pseudo-Herm-SM} \\
  \big\{H, \Sigma_3 \Sigma_1 T \big\} &= 0, \label{anti-PT-SM}
\end{align}
where
\begin{equation}
  \Sigma_\mu = \begin{bmatrix}0 & \sigma_\mu \\ \sigma_\mu & 0
  \end{bmatrix}, \;\;\;\mu = 0, 1, 2, 3.
\end{equation}
An important property of the $\Sigma$ matrices is that $\{\Sigma_i, \Sigma_j\} = 2 \delta_{ij}$ for $i,j = 1,2,3$.

Eq.~\eqref{pseudo-Herm-SM} is an instance of pseudo-Hermiticity \cite{Mostafazadeh2002}.  If a Hamiltonian $H$ is pseudo-Hermitian and non-singular, its eigenvalues are either real or come in complex conjugate pairs.  This is because the uniqueness of the spectrum $\{E_n\}$ implies a set of left eigenvectors $\{\langle \phi_n|\}$ such that $\langle\phi_n|H = \langle\phi_n| E_n$, and the Hermitian conjugates of these are eigenvectors of $H^\dagger$: i.e., $H^\dagger |\phi_n\rangle = E_n^* |\phi_n\rangle$.  Eq.~\eqref{pseudo-Herm-SM} then implies that
\begin{equation}
  H \Big(\Sigma_0 |\phi_n\rangle\Big) = \Sigma_0 H^\dagger |\phi_n\rangle
  = E_n^* \Big(\Sigma_0 |\phi_n\rangle\Big).
\end{equation}
For a more detailed discussion, refer to Ref.~\onlinecite{Mostafazadeh2002}.

We now turn now to the anti-PT symmetry, Eq.~\eqref{anti-PT-SM}.  Consider an eigenstate
\begin{equation}
  H |\psi\rangle = E |\psi\rangle.
\end{equation}
Multiplying this by $\Sigma_1\Sigma_3 T$:
\begin{align}
  \begin{aligned}
    \Sigma_1\Sigma_3 T H |\psi\rangle
    &= E^* \Sigma_1\Sigma_3 T |\psi\rangle \\
    = - \Sigma_3\Sigma_1 T H |\psi\rangle & \\
    = \;\;\, H \Sigma_3 \Sigma_1 T |\psi\rangle &
    \qquad\qquad\qquad\qquad\qquad
    [\textrm{using}\;\;\textrm{Eq.}~\eqref{anti-PT-SM}] \\
    \Rightarrow \quad H \Big(\Sigma_1 \Sigma_3 T |\psi\rangle\Big)
    &= - E^* \Big(\Sigma_1 \Sigma_3 T |\psi\rangle\Big).
  \end{aligned}
\end{align}
Hence, $|\psi'\rangle = \Sigma_1 \Sigma_3 T |\psi\rangle$ is an eigenstate with eigenvalue $-E^*$.  To show that it is orthogonal to $|\psi\rangle$, note that
\begin{equation}
  \Sigma_1 \Sigma_3
  = -i \begin{bmatrix}\sigma_2 & 0 \\ 0 & \sigma_2 \end{bmatrix}.
\end{equation}
Then let
\begin{equation}
  |\psi\rangle
  = \begin{bmatrix} |\varphi_+\rangle \\ |\varphi_-\rangle \end{bmatrix},
  \qquad
  |\psi'\rangle = \Sigma_1 \Sigma_3 T |\psi\rangle
  = -i \begin{bmatrix} \sigma_2T|\varphi_+\rangle \\ \sigma_2T|\varphi_-\rangle \end{bmatrix}.
\end{equation}
Then
\begin{equation}
  \langle \psi | \psi'\rangle = -i
  \Big(\langle \varphi_+ | \sigma_2 T |\varphi_+\rangle
  + \langle \varphi_- | \sigma_2 T |\varphi_-\rangle \Big).
\end{equation}
This vanishes because for any two-component vector $|\varphi\rangle$,
\begin{equation}
  |\varphi\rangle = \begin{bmatrix}a \\ b \end{bmatrix}
  \; \Rightarrow \;
  \langle \varphi | \sigma_2 T |\varphi\rangle = i a^*b^* -i b^*a^* = 0.
\end{equation}

\section{Band Structure of the Non-Hermitian Lattice}

The $k$-space Hamiltonian for the lattice described in the main text is
\begin{align}
  \begin{aligned}
    H &= \begin{bmatrix}\mathcal{W} & \mathcal{V} \\ \mathcal{V} & \mathcal{W}^\dagger \end{bmatrix}, \\
    \mathcal{W}(k) &= \begin{bmatrix}
      m + i\gamma + 2t_2 \sin k_1
      & 2\cos(k_1/2) e^{i\sqrt{3}k_2/6} \\
      2\cos(k_1/2) e^{-i\sqrt{3}k_2/6}
      & -m+i\gamma - 2t_2 \sin k_1
    \end{bmatrix}, \\
    \mathcal{V}(k) &=
    \begin{bmatrix}
      -4t_2 \sin(k_1/2) \cos(\sqrt{3}\,k_2/2)
      & e^{-i\sqrt{3}k_2/3} \\
      e^{i\sqrt{3}k_2/3}
      & 4t_2 \sin(k_1/2) \cos(\sqrt{3}\,k_2/2)
    \end{bmatrix}.
  \end{aligned}
\end{align}
For $m = t_2 = 0$, the eigenenergies are $\{\pm E_1(k), \pm E_2(k)\}$, where
\begin{align}
\begin{aligned}
  E_1 &= \sqrt{3-\gamma^2 + 2\cos(k_1)
    - 2\sqrt{2\cos^2(k_1/2) \left(\cos(\sqrt{3}k_2) + 1 - 2\gamma^2\right)}} \\
  E_2 &= \sqrt{3-\gamma^2 + 2\cos(k_1)
    + 2\sqrt{2\cos^2(k_1/2) \left(\cos(\sqrt{3}k_2) + 1 - 2\gamma^2\right)}}.
\end{aligned}
\end{align}
Thus, for $|\gamma| < 1$, the energies are real inside the region
\begin{equation}
  |k_2| < \frac{\cos^{-1}\left(2\gamma^2-1\right)}{\sqrt{3}},
\end{equation}
as well as along the curves $k_1 = \pm\pi$.  Zero-energy band degeneracy points occur at
\begin{equation}
  K^\pm = \begin{pmatrix} -2 \tau \theta \\ 0
  \end{pmatrix}, \;\;\;\mathrm{where}\;\;
  \begin{cases}
  \cos 2\theta &= -(1+\gamma^2)/2, \\
  \tau &= \pm 1.
  \end{cases}
\end{equation}
The $\tau$ variable describes which degeneracy point we are referring to, similar to the valley index in graphene.

To derive the effective Hamiltonian at the band degeneracy points, we linearize $H(k)$ as described in the main text.  It is convenient to define
\begin{equation}
  k = \begin{pmatrix} -2\tau\theta + q_1 \\ \sqrt{3} \,\tilde{q}_2
  \end{pmatrix},
\end{equation}
where $\tilde{q}_2 = q_2/\sqrt{3}$.  Consider an eigenstate of the form
\begin{equation}
  |\psi\rangle = \begin{bmatrix} |\varphi_+\rangle \\ |\varphi_-\rangle
  \end{bmatrix}, \label{two-components}
\end{equation}
where $|\varphi_\pm\rangle$ are two-component vectors, with energy $E$.  Plugging this into the Schr\"odinger equation gives
\begin{align}
  \Big[\mathcal{W}_0 + \mathcal{W}_1(q)\Big] |\varphi_+\rangle
  + \Big(\sigma_1 + \sigma_2 \tilde{q}_2 + 4\tau t_2 \sin\theta \sigma_3\Big)
  |\varphi_-\rangle
  &= E |\varphi_+\rangle \label{master1} \\
  \Big(\sigma_1 + \sigma_2 \tilde{q}_2 + 4\tau t_2\sin\theta\sigma_3 \Big)
  |\varphi_+\rangle
  + \Big[\mathcal{W}_0^\dagger + \mathcal{W}_1(q)\Big] |\varphi_-\rangle
  &= E |\varphi_-\rangle, \label{master2}
\end{align}
where
\begin{align}
  \begin{aligned}
    \mathcal{W}_0 &=
    i \gamma \sigma_0 + \big[m - 2\tau t_2 \sin(2\theta)\big] \sigma_3
    + 2 \cos\theta \, \sigma_1 \\
    \mathcal{W}_1 &= \tau \sin\theta \, \sigma_1 \, q_1
    - \cos\theta \, \sigma_2\, \tilde{q}_2.
  \end{aligned}
  \label{W01}
\end{align}
Multiplying Eq.~\eqref{master1} by $(\sigma_1+\sigma_2\tilde{q}_2 + 4 \tau t_2 \sin\theta \sigma_3)$, and assuming that $t_2 \sim O(q)$, gives
\begin{equation}
  |\varphi_-\rangle
  = (\sigma_1 + \sigma_2 \tilde{q}_2 + 4 \tau t_2 \sin\theta \sigma_3)
  (E - \mathcal{W}_0 - \mathcal{W}_1) |\varphi_+\rangle + O(q^2).
  \label{working1}
\end{equation}
Plugging this into Eq.~\eqref{master2} gives
\begin{equation}
  \big(\sigma_1 + \sigma_2 \tilde{q}_2 + 4 \tau t_2 \sin\theta \sigma_3\big)
  \, |\varphi_+\rangle =
  \big(E - \mathcal{W}_0^\dagger - \mathcal{W}_1\big)
  \big(\sigma_1 + \sigma_2 \tilde{q}_2 + 4 \tau t_2 \sin\theta \sigma_3\big)
  \big(E - \mathcal{W}_0 - \mathcal{W}_1\big) |\varphi_+\rangle + O(q^2).
\end{equation}
Now we further assume that $E, m \sim O(q)$, and expand the polynomials on the right-hand side to first order.  (Note that $\mathcal{W}_0$ is zeroth-order in $q$ and $\mathcal{W}_1$ is first-order.)  We obtain
\begin{multline}
  \Big[\big(\mathcal{W}_0^\dagger \sigma_1 \mathcal{W}_0 - \sigma_1 \big)
  + \big(\mathcal{W}_0^\dagger \sigma_1 \mathcal{W}_1
  + \mathcal{W}_1 \sigma_1 \mathcal{W}_0\big) \\
  + \big(\mathcal{W}_0^\dagger \sigma_2 \mathcal{W}_0 - \sigma_2\big)
  \tilde{q}_2
  + 4\tau t_2\sin\theta \big(\mathcal{W}_0^\dagger \sigma_3\mathcal{W}_0 - \sigma_3\big)
  \Big] \, |\varphi_+\rangle
  = E \Big(\sigma_1 \mathcal{W}_0 + \mathcal{W}_0^\dagger \sigma_1 \Big)
  |\varphi_+\rangle + O(q^2). \label{schrodi}
\end{multline}
Using Eq.~\eqref{W01}, we have
\begin{align}
  \begin{aligned}
    \mathcal{W}_0^\dagger \sigma_1 \mathcal{W}_0 - \sigma_1
    &= 4\cos\theta\,\big(m
    - 2\tau t_2 \sin2\theta\big)\, \sigma_3^+
    + O(q^2) \\
    \mathcal{W}_0^\dagger \sigma_1 \mathcal{W}_1
    + \mathcal{W}_1 \sigma_1 \mathcal{W}_0
    &= 4 \cos\theta\left( \tau \sin\theta\, \sigma_1 q_1 -
    \cos\theta\, \sigma_2^+ \, \tilde{q}_2\right) + O(q^2) \\
    \mathcal{W}_0^\dagger \sigma_2 \mathcal{W}_0 - \sigma_2
    &= -8\cos^2\theta\; \sigma_2^+ + O(q)\\
    \mathcal{W}_0^\dagger \sigma_3\mathcal{W}_0 - \sigma_3
    &= -8\cos^2\theta\; \sigma_3^+ + O(q) \\
    \sigma_1 \mathcal{W}_0 + \mathcal{W}_0^\dagger \sigma_1
    &= 4\cos\theta \, \sigma_0.
  \end{aligned}
\end{align}
where we define the $\gamma$-dependent modified Pauli matrices
\begin{align}
  \sigma_2^\pm &= \sigma_2 \pm \frac{\gamma}{2\cos\theta}\sigma_3 \\
  \sigma_3^\pm &= \sigma_3 \mp \frac{\gamma}{2\cos\theta}\sigma_2.
\end{align}
Eq.~\eqref{schrodi} simplifies to
\begin{equation}
  \big(\,\tau \alpha_1 q_1 + \alpha_2^+ q_2 + \beta_+\,\big)
  |\varphi_+\rangle = E \, |\varphi_+\rangle + O(q^2),
\end{equation}
where we have set $q_2 = \sqrt{3}\,\tilde{q}_2$ and
\begin{align}
  \alpha_1 &= \sin\theta \, \sigma_1 \\
  \alpha_2^\pm
  &= - \sqrt{3} \cos\theta \, \sigma_2^{\pm} \\
  \beta_\pm &= (m - 6\tau t_2 \sin2\theta)\, \sigma_3^\pm
\end{align}
We can repeat the above procedure, solving for the lower two components $\varphi_-$.  The resulting linearized Schr\"odinger equation is similar to Eq.~\eqref{schrodi}, but with $\mathcal{W}_0$ and $\mathcal{W}_0^\dagger$ swapped, which is the same as swapping $\gamma \leftrightarrow -\gamma$.  Hence,
\begin{equation}
  \big(\tau \alpha_1 q_1 + \alpha^\pm_2 q_2
    + \beta_\pm\big) |\varphi_\pm\rangle = E \, |\varphi_\pm\rangle + O(q^2).
    \label{distorted-Dirac-SM}
\end{equation}
To simplify this further, define the $\gamma$-dependent unitary operator
\begin{equation}
  U_\pm = \exp\left(\pm \frac{i\phi}{2}\sigma_1\right) \;\;\mathrm{where}\;\;
  \phi = \cos^{-1}(2\cos\theta) = \sin^{-1}\gamma.
\end{equation}
Then
\begin{align}
  U_\pm \, \sigma_2^\pm \, U_\mp &= \frac{1}{2\cos\theta}\, \sigma_2 \\
  U_\pm \, \sigma_3^\pm \, U_\mp &= \frac{1}{2\cos\theta}\, \sigma_3.
\end{align}
Now Eq.~\eqref{distorted-Dirac-SM} can be transformed as follows:
\begin{align}
  \frac{\sqrt{3}}{2} \Big(\tau \eta_\gamma \sigma_1 q_1 - \sigma_2 q_2
  + M(\gamma) \sigma_3 \Big) U_\pm |\varphi_\pm\rangle &= E \, U_\pm |\varphi_\pm\rangle + O(q^2) \label{Dirac-SM} \\
  \eta_\gamma &= 2\sin\theta/\sqrt{3} \\
  M(\gamma) &=
  \frac{m - 6\tau t_2 \sin2\theta}{2\cos\theta}.
\end{align}

\section{Massless Dirac Cones and Non-Hermitian Landau Levels}

For $m = t_2 = 0$, the band structure simplifies considerably.  In this case, the Hamiltonian obeys the symmetries
\begin{align}
  \Big\{H(k)\; ,\; \Sigma_3 \Big\} &= 0, \label{PH} \\
  \Big[H(k) \;,\; \Sigma_1 T \Big] &= 0. \label{PT}
\end{align}
These are, respectively, a particle-hole symmetry and a PT (parity/time-reversal) symmetry.  Eqs.~\eqref{PH} and \eqref{PT} together imply the anti-PT symmetry $\{H(k), \Sigma_3 \Sigma_1 T \} = 0$, but the reverse is not necessarily true.

The particle-hole symmetry \eqref{PH} implies that if $H |\psi\rangle = E |\psi\rangle$, then $H \Sigma_3 |\psi\rangle = - E \Sigma_3 |\psi\rangle$.  Moreover, if $|\psi\rangle$ does not spontaneously break the PT symmetry \eqref{PT}, then $E = E^*$ and $\Sigma_1 \mathcal{T} |\psi\rangle = \exp(i\phi) |\psi\rangle$ for some $\phi$.  Now consider an eigenstate of the form \eqref{two-components}.  If PT is unbroken, and the $\Sigma_1 T$ eigenvalue is $\exp(i\phi)$, then
\begin{equation}
  \sigma_1 T |\varphi_+\rangle = e^{i\phi} |\varphi_-\rangle
  \;\;\textrm{and}\;\;
  \sigma_1 T |\varphi_-\rangle = e^{i\phi} |\varphi_+\rangle.
\end{equation}
By setting the global phase degree of freedom for $|\psi\rangle$, we can always choose $\phi = 0$, so that
\begin{equation}
  |\varphi_-\rangle = \sigma_1 T \, |\varphi_+\rangle.
\end{equation}
Thus, it is really only necessary to keep track of $|\varphi_+\rangle$.

For the ungapped Dirac cones, Eq.~\eqref{Dirac-SM} simplifies to
\begin{equation}
  v\; \Big(\tau \eta_\gamma \sigma_1 q_1 - \sigma_2 q_2 \Big)
  U_+ |\varphi_+\rangle \simeq E \, U_+ |\varphi_+\rangle
  \label{Dirac-cone-SM}
\end{equation}
where $v = \sqrt{3}/2$, $U_+ = \exp[i(\phi/2)\sigma_1]$, $\phi = \sin^{-1}\gamma$, and
\begin{equation}
  \eta_\gamma = 2\sin\theta/\sqrt{3}.
  \label{eta-SM}
\end{equation}
Note that $q = k - K^\tau(\gamma)$ is the $k$-space displacement relative to the Dirac point at gain/loss parameter $\gamma$.

Now consider slow spatial variations in $\gamma$.  The microscopic wavefunction can be described by the ansatz
\begin{equation}
  \psi(r) = \varphi_A(r) u_A(r) + \varphi_B(r) u_B(r),
\end{equation}
where $u_{A/B}(r)$ are the Bloch wavefunctions at some reference Dirac point, $K^\tau_1(\theta_0) = -2\tau\theta_0$.  The slowly-varying envelope fields $\varphi_{A/B}(r)$ form a two-component field; let us define
\begin{equation}
  \Phi(r) = U_+ \begin{bmatrix}\varphi_A(r) \\ \varphi_B(r) \end{bmatrix}
  = \begin{bmatrix}\Phi_A(r) \\ \Phi_B(r) \end{bmatrix},
  \label{Phir-SM}
\end{equation}
where $U_+$ may vary in space via its dependence on $\gamma$.  Eq.~\eqref{Dirac-cone-SM} translates into the wave equation
\begin{equation}
  v\; \Big[\tau \eta_\gamma \sigma_1 (-i\partial_1 + A_1)
    + i \sigma_2 \partial_2 \Big] \Phi(r) \simeq E \, \Phi(r),
  \label{Dirac-wave-equation-SM}
\end{equation}
where $A_1$ is a gauge field that characterizes the spatial variation of $\gamma$.  There is no $x_2$ gauge field component since the Dirac cones only move along $k_1$ when $\gamma$ is varied.  When $\gamma$ is spatially uniform, we have a plane wave solution
\begin{equation}
  \Phi(r) = e^{i\kappa \cdot r} \Phi,\qquad
  v\; \Big[\tau \eta_\gamma \sigma_1 (\kappa_1 + A_1)
    - \sigma_2 \kappa_2 \Big] \Phi \simeq E \, \Phi,
\end{equation}
where $\kappa_1 = k - K^\tau_1(0)$ is the $k$-vector component relative to the reference Dirac point.  Comparing to Eq.~\eqref{Dirac-cone-SM} gives
\begin{equation}
  A_1 = - K_1^\tau(\gamma) + K_1^\tau(0) = 2\tau (\theta - \theta_0).
\end{equation}

As in the main text, let us now suppose $\gamma$ varies only along $x_2$, so that there is a conserved momentum $\kappa_1$.  Moreover, we suppose that the variation of $\gamma$ is such that $\theta = \theta_0 + \beta x_2$, where $\beta x_2 \ll 1$.  From Eq.~\eqref{eta-SM},
\begin{equation}
  \eta_\gamma = \frac{2}{\sqrt{3}}\Big(\sin\theta_0
  + \beta x_2 \cos\theta_0\Big) + O(\beta^2x_2^2).
\end{equation}
The first term inside the square parentheses of  Eq.~\eqref{Dirac-wave-equation-SM} becomes
\begin{equation}
  \tau \eta_\gamma \sigma_1 [\kappa_1 + 2\tau(\theta-\theta(0))] \approx
  \frac{\Omega}{\sqrt{2}} (x_2 - \mu)\, \sigma_1
\end{equation}
where
\begin{align}
  \Omega &= \sqrt{\frac{8}{3}}
  \left(2\sin\theta_0+\kappa_1\tau\cos\theta_0\right) \beta
  \label{Omega-SM}\\
  \mu &= - \frac{\kappa_1\tau\sin\theta_0}{\Omega \sqrt{3/8}}.
\end{align}
Eq.~\eqref{Dirac-wave-equation-SM} then reduces to
\begin{align}
  v\; \left[\frac{\Omega}{\sqrt{2}}(x_2-\mu) + \partial_2 \right]
  \Phi_B(x_2) \simeq E \, \Phi_A(x_2) \label{Jackiw1} \\
  v\; \left[\frac{\Omega}{\sqrt{2}}(x_2-\mu) - \partial_2 \right]
  \Phi_A(x_2) \simeq E \, \Phi_B(x_2) \label{Jackiw2}
\end{align}
This has two types of solutions.  The first are the zero modes, which arise when Eqs.~\eqref{Jackiw1} and \eqref{Jackiw1} are both zero:
\begin{equation}
  E = 0, \quad
  \Phi_{A/B}(x_2) \sim \exp\left[\int^{x_2} dx_2' \;
    \frac{\Omega}{\sqrt{2}}(x_2'-\mu)\right] \; \Phi_{A/B}^{(0)},
  \quad
  \Phi_{B/A}(r) = 0.
\end{equation}
The envelope fields are then given by $\varphi(r) = U_+^\dagger \Phi(r)$ [Eq.~\eqref{Phir-SM}].  Note that because of the $U_+^\dagger$ operator, the zero modes are not sublattice polarized.

The second type of solution is the non-zero modes.  These are obtained by combining Eqs.~\eqref{Jackiw1}--\eqref{Jackiw2}:
\begin{equation}
  a a^\dagger\, \Phi_A = \frac{E^2}{\sqrt{2}\Omega v^2}\Phi_A,
  \;\;\;\mathrm{where}\;\;\;
  a = \frac{1}{\left(\sqrt{2}\Omega\right)^{1/2}}
  \left[\frac{\Omega}{\sqrt{2}}(x_2 - \mu) + \partial_2\right].
\end{equation}
Note that $[a, a^\dagger]=1$.  The resulting spectrum is
\begin{equation}
  E = \pm v\sqrt{\sqrt{2}\Omega n}
  \;\;\;n=1,2,3\dots
\end{equation}
Because $\Omega$ depends on $\kappa_1$ [see Eq.~\eqref{Omega-SM}], these bands are not perfectly flat.

\clearpage
\end{widetext}

\end{document}